\documentstyle[times,pramana,epsfig,floats]{ias}
\def\chio{\tilde\chi{^0_1}}

\def\mt{\mbox{$m_{\rm \tilde{\rm t}_1}$}}
\def\mchi{m_{\chio}}
\def\cost{\mbox{$\cos\theta_{\rm \tilde t}$}}
\newcommand{\neu}{\tilde{\chi}^0}
\newcommand{\mst}{m_{\rm \tilde{t}_1}}
\newcommand{\mneu}[1]{m_{\tilde{\chi}^0_{#1}}}

\begin{document}
\newcommand{\stp}{\tilde t_1}
\def\chio{\textstyle\raise.4ex\hbox{$\textstyle\tilde\chi{^0_1}$}}

\begin{titlepage}

\thispagestyle{empty}
\def\thefootnote{\fnsymbol{footnote}}       % symbols for footnotes

\begin{center}
\mbox{ }

\end{center}
\begin{flushright}
\vspace* {-3.0cm}
\Large
\mbox{\hspace{10.2cm} physics/0609017} \\
\end{flushright}
\begin{center}
\vskip 1.0cm
{\Huge\bf
Scalar Top Study: 
}
\vspace{2mm}

{\Huge\bf
Detector Optimization
}
\vskip 1cm
{\LARGE\bf C. Milst\'ene$^1$, A. Sopczak$^2$}
\smallskip

\Large $^1$Fermilab, USA; $^2$Lancaster University, UK

\vskip 2.5cm
\centerline{\Large \bf Abstract}
\end{center}

\vskip 3.5cm
\hspace*{-0.5cm}
\begin{picture}(0.001,0.001)(0,0)
\put(,0){
\begin{minipage}{\textwidth}
\Large
\renewcommand{\baselinestretch} {1.2}
A vertex detector concept of the Linear Collider Flavour Identification 
(LCFI) collaboration, which studies pixel detectors for heavy quark flavour identification,
has been implemented in simulations for c-quark tagging in scalar top studies.
The production and decay of scalar top quarks (stops) is particularly interesting
for the development of the vertex detector as only two c-quarks
and missing energy (from undetected neutralinos) are produced for light stops.
Previous studies investigated the vertex detector design in
scenarios with large mass differences between stop and neutralino, corresponding
to large visible energy in the detector. In this study we investigate the tagging
performance dependence on the vertex detector design in a scenario with small visible energy
for the International Linear Collider (ILC).
\renewcommand{\baselinestretch} {1.}

\normalsize
\vspace{2.5cm}
\begin{center}
{\sl \large
Presented at the 2006 International Linear Collider Workshop - Bangalore, 
India, \\
to be published in the proceedings.
\vspace{-6cm}
}
\end{center}
\end{minipage}
}
\end{picture}
\vfill

\end{titlepage}

\newpage
\thispagestyle{empty}
\mbox{ }
\newpage
\setcounter{page}{1}

\mark{{Linear Collider Workshop}{C. Milst\'ene and A. Sopczak}}
\title{Scalar Top Study: Detector Optimization}

\author{C. Milst\'ene$^1$ and A. Sopczak$^2$}
\address{$^1$Fermilab, USA; $^2$Lancaster University, UK}

\abstract{
A vertex detector concept of the Linear Collider Flavour Identification 
(LCFI) collaboration, which studies pixel detectors for heavy quark flavour identification,
has been implemented in simulations for c-quark tagging in scalar top studies.
The production and decay of scalar top quarks (stops) is particularly interesting
for the development of the vertex detector as only two c-quarks
and missing energy (from undetected neutralinos) are produced for light stops.
Previous studies investigated the vertex detector design in
scenarios with large mass differences between stop and neutralino, corresponding
to large visible energy in the detector. In this study we investigate the tagging
performance dependence on the vertex detector design in a scenario with small visible energy
for the International Linear Collider (ILC).
}

\maketitle
\section{Introduction}
The development of a vertex detector for a Linear Collider 
is an important and challenging enterprise.
A key aspect is the distance of the innermost layer to the interaction point,
which is related to radiation hardness and beam background. Another key aspect
is the number of radiation lengths the particles go through, since it determines
the multiple scattering which affects the vertex reconstruction.

The optimization of the vertex detector tagging performance 
is of great importance for studies of physics processes.
While mostly at previous and current accelerators (e.g. SLC, LEP, Tevatron) b-quark 
tagging has revolutionized many searches and measurements, c-quark tagging will 
be very important at a future Linear Collider, for example, in studies of Supersymmetric 
dark matter~\cite{Carena:2005gc}.
Therefore, c-quark tagging could be a benchmark for vertex detector developments. 
The scalar top production and decay process, and the implemented vertex detector geometry 
are shown before~\cite{det_snowmass05}.

The analysis for a large mass difference with the SPS-5 parameter point (ISAJET)
$\mt = 220.7$~GeV, $\mchi = 120.0$ GeV and $\cost = 0.5377$ was previously 
performed~\cite{andre_lcws04}.
For 25\% (12\%) efficiency 3800 (1800) signal events and 5400 (170) background events 
without c-quark tagging were obtained, 
while the background was reduced to 2300 (68) events with c-quark tagging.

The vertex detector radiation length was varied between single thickness (TESLA TDR)
and double thickness. In addition, the number of vertex detector layers was varied between
5 layers (innermost layer at 1.5 cm as in the TESLA TDR) and 4 layers (innermost layer at 
2.6 cm).
For SPS-5 parameters the following numbers of background events remain~\cite{andre_lcws04}:
\begin{center}
\vspace*{-1mm}
\begin{tabular}{l|c|cc}
Thickness& layers & 12\% signal efficiency~~~~ &25\% signal efficiency \\
\hline
    Single &  5 (4)  &   68 (82) & 2300 (2681) \\
    Double &  5 (4)  &   69 (92) & 2332 (2765) 
\end{tabular}
\vspace*{-6mm}
\end{center}
A significant larger number of background events was expected if the first layer of the 
vertex detector is removed. The distance of the first layer to the interaction point is also
an important aspect from the accelerator physics (beam delivery) perspective. The interplay between
the beam delivery and vertex detector design in regard to critical tolerances like hardware damage of the 
first layer and occupancy (unable to use the data of the first layer) due to beam background goes
beyond the scope of this study and will be addressed in the future.

For large visible energy (large mass difference) no significant increase in the expected background was 
observed for doubling the thickness of the vertex detector 
layers~\cite{andre_lcws04,susy05}. In this study the effect of the vertex 
detector design for events with smaller visible energy in the detector is addressed.

\section{Signal and Background Simulations}

The production of simulated light stops at a 500 GeV Linear Collider is
analyzed using high luminosity ${\cal L} = 500~{\rm fb}^{-1}$.
The signature for stop pair production at an $\rm e^+e^-$ collider is two 
charm jets and large missing energy:
\begin{equation}
\rm e^+e^- \to \tilde{t}_1 \, \bar{\tilde{t}}_1 \to c \, \neu_1 \, \bar{c} \, \neu_1.
\end{equation}
For small $\Delta m = \mst - \mneu{1}$, the jets are
relatively soft and separation from backgrounds is very challenging.
Backgrounds arising from various Standard Model processes can have
cross-sections that are several orders of magnitude larger than the signal.
Thus, it is necessary to study this process with a realistic detector simulation.
Signal and background events are generated with {\sc Pythia 6.129}~\cite{Sjostrand:2000wi}, 
including a scalar top signal generation~\cite{sopczak-stop-gen} 
previously used in Ref.~\cite{as_stopsnew}.  The detector simulation is
based on the fast simulation {\sc Simdet}~\cite{Pohl:2002vk}, 
describing a typical ILC detector. Good agreement in comparisons with
{\sc SGV}~\cite{berggren} detector simulations was obtained~\cite{andre_lcws04,susy05}.

Cross-sections for the signal process and the relevant backgrounds 
have been computed with code used in Ref.~\cite{slep} and 
by {\sc Grace 2.0} \cite{grace}, with cross-checks to {\sc CompHep 4.4}~\cite{comphep}. 
A minimal transverse momentum cut, $p_{\rm t} > 5$ GeV, is applied for the two-photon 
background, to avoid the  infrared divergence. Details of the event selection are given in
Ref.~\cite{Carena:2005gc}.

The c-tagging with the LCFI detector is based on the vertex identification and a neural network 
application~\cite{kuhl}. The vertex identification considered three cases for
each jet independently:
\begin{itemize}
\item [a)] only a primary vertex. In this case, the two tracks with the largest 
           separation in the $r$-$\phi$ plane are considered and  for these tracks the neural network 
           variables include, the impact parameter and its significance 
           (impact parameter divided by uncertainty)
           both in the $r$-$\phi$ plane and in the $z$-direction, their momenta, and the 
           joined probability in $r$-$\phi$ plane and $z$ direction.
\item [b)] one secondary vertex. In addition to the previous variables, the decay length significance
           of the secondary vertex, the multiplicity and momenta of all associated tracks, and
           the $P_t$ corrected mass of the secondary vertex (corrected for neutral hadrons
           and neutrinos), the $P_t$ of the decay products perpendicular to the flight direction
           between primary and secondary vertex, and the joint probability in $r$-$\phi$ and
           $z$-direction.
\item [c)] more than one secondary vertex. Two secondary vertices are considered,
           where the tracks are assigned to the vertex closest to the primary vertex, and
           the neural network input variables are defined as in case b).
\end{itemize}
The neural network is tuned with 255,000 simulated signal and 240,000 $\rm W e \nu$ background 
events. The signal events are a combination of all simulated signal events for the 
scalar top mass range between 120 to 220 GeV and for $\Delta m=5,10$ and~15~GeV. 

After a preselection which substantially reduces the background while keeping about
70\% of the signal, six sequential cuts are applied: number of jets,
missing energy, acollinearity, thrust angle, transverse momentum, and the 
jet-jet invariant mass and c-tagging~\cite{Carena:2005gc}.
The background consists of the following processes $\rm W^+W^-$, $\rm ZZ$, $\rm W e\nu$, 
$\rm e e Z$, $\rm q \bar{q} (\,q \neq t)$, $\rm t \bar{t}$, and two-photon.
After all cuts, the total background of about 5680 events is dominated by about 5044 $\rm W e\nu$ events~\cite{Carena:2005gc}. 
A  scalar top signal of 120 GeV has been simulated with a neutralino mass of 110 GeV.
The selection efficiency is about 20\%\footnote{The retuning of the c-tagging neural 
network increased the selection efficiency from 19\%~\cite{Carena:2005gc} to 20\% 
and the $\rm We\nu$ background from 5044~\cite{Carena:2005gc} to $5322\pm280$ events.}
and 11,500 signal events are expected for a standard LCFI vertex detector configuration as given in the 
TESLA TDR.

\section{Varying the Vertex Detector Design}

\begin{figure}[htbp]
\vspace*{-0.2cm}
\begin{center}
\includegraphics[width=\textwidth,height=6cm]{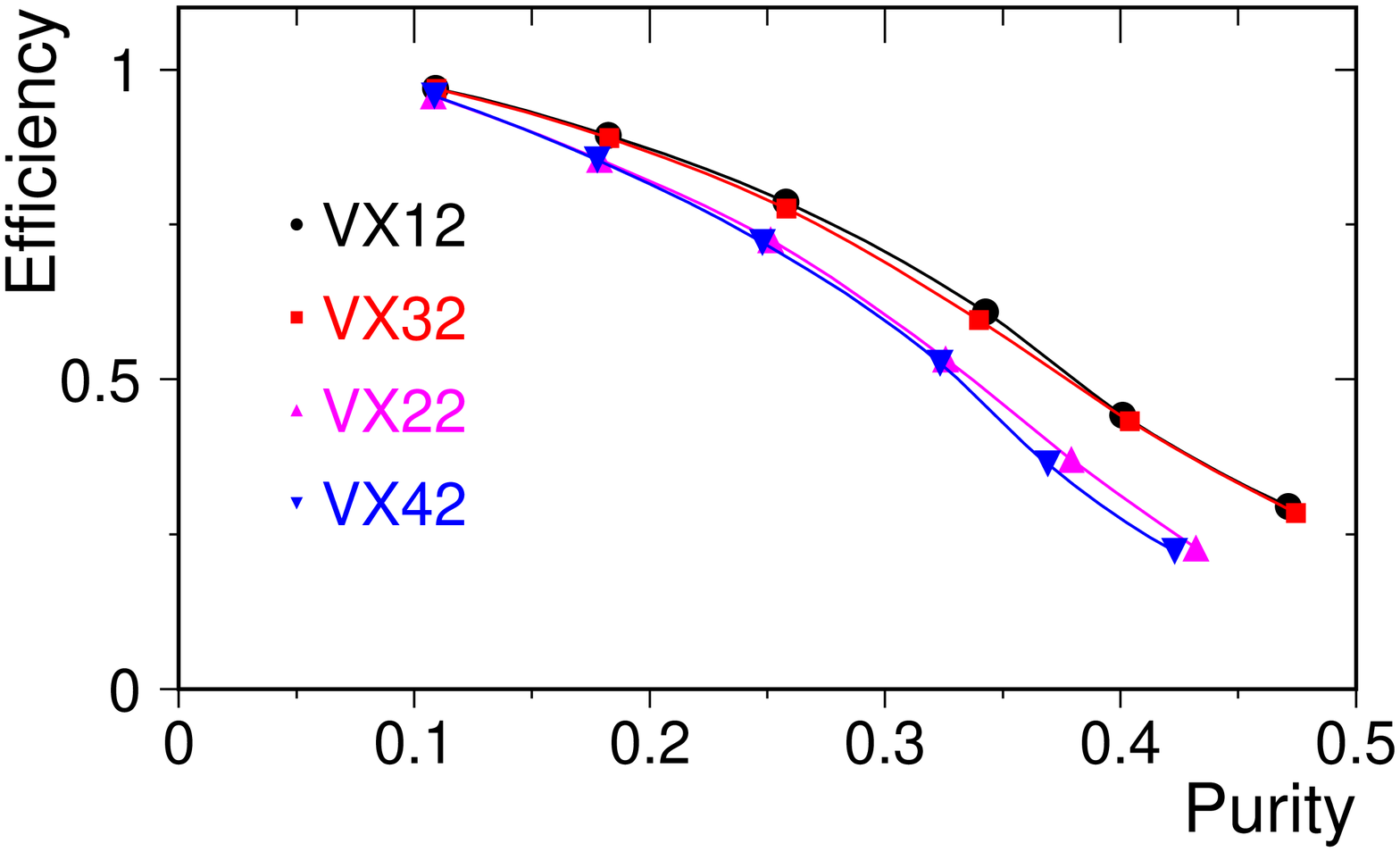}
\end{center}
\vspace*{-0.3cm}
\caption{Left: scalar top c-tagging efficiency and purity with $\rm W e\nu$ background 
for different detector designs. The VX$_{12}$ curve is for a detector design with 5 layers 
(innermost at 15~mm) and single density, curve VX$_{22}$ is for a detector design with 4
 layers (innermost at 26~mm). Curves VX$_{32}$ and VX$_{42}$ are for double density
(0.128\% $X_0$ radiation length per layer) with 4 and 5 layers, respectively.
Right: expected signal efficiency and number of $\rm We\nu$ background events for the four 
detector designs. The statistical uncertainties are also given.
} \label{fig:design}
\vspace*{-0.6cm}
\end{figure}

This study of the vertex detector design is based on 50,000 simulated 120 GeV signal and 210,000 
$\rm W e \nu$ background events for each detector design.
After preselection 29,842 signal and 53,314 $\rm We\nu$ events are selected, corresponding to
34,318 and 779,450 events per 500~fb$^{-1}$, respectively.
This preselection signal efficiency of 59.7\% does not depend on the vertex detector design.
Four detector designs are compared:
\begin{itemize}
\item[VX$_{12}$:] the TESLA TDR design with 5 layers and single (0.064\% $X_0$ radiation length
                 per layer).
\item[VX$_{22}$:] 4 layers (the innermost layer removed). This scenario could for example occur if 
                 the vertex detector is exposed to a large dose of machine background from the 
                 accelerator.
                 The optimization of the radius of the innermost layer is an important
                 aspect in the design of a vertex detector for a Linear Collider.
\item[VX$_{32}$:] 5 layers and double material thickness (0.128\% $X_0$ radiation length per layer).
                 As the rigidity of the sensitive elements and the support structure is another
                 important aspect in the detector design, the material budget has to be taken into account.
\item[VX$_{42}$:] 4 layers (the innermost layer removed) and double thickness 
                 (0.128\% $X_0$ radiation length per layer). 
\end{itemize}
The c-tagging efficiency per event is normalized to the number of signal events 
after the preselection and requiring two jets. At least one c-tagged jet is required 
and the efficiency is given in Fig.~\ref{fig:design} as a function of the purity,
where purity is defined as the ratio of the number of simulated signal events after the c-tagging  
to all c-tagged events assuming the same luminosity for signal and background.
The different purities are obtained by varying the cut on the c-tagging neural network variable.
The effect of the detector design variation increases with increasing purity (harder c-tagging neural network cut).
For the second set of points in the plot with purities about 18\% and c-tagging efficiencies between 85\% and 90\%, 
the variation of the signal efficiency and the number of $\rm We\nu$ background events is given in the 
table after all selection cuts.
\begin{center}
\begin{tabular}{l|c|c|c|c}
Thickness& layers & signal efficiency (in \%)&$\rm We\nu/210k$ &$\rm We\nu/500~fb^{-1}$ \\
\hline
    Single &  5 (4)  &   20.46 (19.67)& 364 (369)&5322 (5396)  \\
    Double &  5 (4)  &   20.32 (19.52)& 366 (385)&5352 (5630)  \\ 
           &  &   $\pm 0.18$         &  $\pm 19$     & $\pm 280$ 
\end{tabular}
\end{center}

\section{Results}
\vspace*{-0.2cm}
These results for small visible energy ($\Delta m = 10$~GeV) lead to the same 
observation as for large visible energy (SPS-5 scenario with $\Delta m = 100.7$~GeV).
The radius of the innermost layer of the vertex detector has a large effect on
the c-quark tagging performance.
Curves VX$_{12}$ and VX$_{22}$ of Fig.~\ref{fig:design} show the performance for radii 15 and 
26~mm, respectively. 
There is no significant effect on the c-quark tagging performance from doubling the material 
budget (e.g. curves VX$_{12}$ and VX$_{32}$ of Fig.~\ref{fig:design}). 
The increase of multiple scattering is not significant between single and double 
thickness (0.128\% $X_0$ radiation length per layer).

In order to quantitatively estimate the multiple scattering effect, the number of tracks
per signal event  and the visible energy have been determined. The minimum visible energy
per event is about 10~GeV and the maximum number of tracks is about 20, therefore the minimum
track energy is about 0.5~GeV. The analytical calculation of the multiple scattering angle 
is given by $ \theta\approx 13.6/P\cdot \sqrt{x/X_0},$ 
where the track momentum $P$ is given in MeV.
The displacement at the interaction point is $d\approx R\theta$, where $R$ is the radius
of the innermost layer of the vertex detector. For $P=500$~MeV, $x/X_0=0.128\%$ and $R=15$~mm, 
$d=15\mu$m. 
This small value compared to the flight distance of charm mesons explains the insignificant effect on
the c-quark tagging from the multiple scattering increase by doubling the vertex detector layer 
thickness.

\section{Conclusions}

In conclusion, the studies with a small visible energy signal lead to the same results 
as in the previous study for large visible energy regarding the vertex detector design.
A strong dependence on the detector performance on the radius of the innermost 
vertex detector layer has been observed, while doubling the material thickness
has no significant effect on the c-quark tagging performance in scalar 
top studies at the ILC.
The optimization of the radius of the innermost vertex detector layer will have
to take into account the accelerator background which depends on the machine optics
and the collimation system.

\section*{Acknowledgements}
AS\,would\,like\,to\,thank\,the\,organizers\,of\,the\,workshop\,for\,making\,this\,presentation\,possible.

\end{document}